\documentclass[12pt]{article}
\usepackage{epsf,amsmath,amssymb,amsfonts}

\renewcommand{\Re}{\operatorname{Re}}

\newcommand{\Tr}{\operatorname{Tr}}
\newcommand{\q}{\boldsymbol q}
\newcommand{\p}{\boldsymbol p}
\newcommand{\ve}{\varepsilon}

\begin{document}
~\vspace{3.0cm}

\centerline{\LARGE Leading off-diagonal approximation for the}
\vspace{0.3cm}
\centerline{\LARGE spectral form factor for uniformly hyperbolic systems}
\vspace{2.0cm}
\centerline{\large Martin Sieber\footnote{E-mail: 
m.sieber@bristol.ac.uk}}
\vspace{0.5cm}

\centerline{School of Mathematics, University of Bristol,
Bristol BS8\,1TW, UK}

\vspace{5.0cm}
\centerline{\bf Abstract}
\vspace{0.5cm}
We consider the semiclassical approximation to the
spectral form factor $K(\tau)$ for two-dimensional
uniformly hyperbolic systems, and derive the first
off-diagonal correction for small $\tau$. The result
agrees with the $\tau^2$-term of the form factor for
the GOE random matrix ensemble.
\vspace{2.5cm}

\noindent PACS numbers: \\
\noindent 03.65.Sq ~ Semiclassical theories and applications. \\
\noindent 05.45.Mt ~ Semiclassical chaos (``quantum chaos'').

\newpage
A fundamental characteristic of quantum systems that are
chaotic in their classical limit is universality. It is
observed that diverse systems behave identically when
statistics of energy levels or wave functions are considered,
provided that they have the same symmetries.
These universal statistics agree with those of random matrix
theory, i.e.\ with the statistics of eigenvalues and 
eigenvectors of large random matrices \cite{BGS84}.
Support for this random
matrix hypothesis comes from a large number of numerical
and experimental investigations which have been carried
out on a great variety of systems \cite{Haa92}. However, it remains
an open question to understand the origin of this universality,
and its relation to the underlying classical dynamics. 

One theoretical approach by which such an understanding may be
attempted is the semiclassical method. Semiclassical
approximations are asymptotically valid in the limit
$\hbar \rightarrow 0$ where universality is expected to hold.
Moreover, they directly connect quantum properties with
properties of the corresponding chaotic classical system. They
have been applied in particular to statistical distributions
of the energy levels which are bilinear in the density
of states, one example being the spectral form factor $K(\tau)$.
One of the successes of the semiclassical approach has been
to show that the spectral statistics do indeed agree with the
random matrix statistics in the limit of long-range
correlations; specifically the correct leading order behaviour
of $K(\tau)$ as $\tau \rightarrow 0$ has been derived \cite{Ber85}.

An extension of this result requires knowledge of
correlations between different periodic orbits \cite{ADDKKSS93}.
The relevant mechanisms by which periodic orbits are correlated
have to be identified, and the contributions of correlated
orbits to the spectral form factor have to be evaluated.
Based on an analogy with disordered systems \cite{WLS99}
and with diffractive corrections \cite{Sie00}
it has been suggested that the next term in the expansion
of $K(\tau)$ for small $\tau$ originates from
`two-loop orbits': orbits that have
a self-intersection with small crossing angle and neighbouring
orbits without self-intersection \cite{SR01}. There is
strong numerical evidence that in systems with time-reversal
symmetry these orbit pairs indeed yield the next-order-term in agreement
with the expectation based on random matrix theory. 

In the following we present a derivation of the next to leading
order term in the expansion of the spectral form factor for small
$\tau$. We evaluate analytically the contributions of the
two-loop orbits to the form factor for uniformly hyperbolic systems
with time-reversal symmetry, and we show that the result indeed
agrees with random matrix theory. The calculation makes clear the
properties of classical trajectories which are responsible for the
universal result.

We consider the spectral form factor, which is defined as the
Fourier transform of the two-point correlation function
of the density of states
\begin{equation} \label{kdef}
K(\tau) = \int_{-\infty}^{\infty} \! \frac{d \eta}{\bar{d}(E)}
\left\langle d_{\text{o}}\!\left( E + \frac{\eta}{2} \right)
             d_{\text{o}}\!\left( E - \frac{\eta}{2} \right)
\right\rangle_E
e^{2 \pi i \eta \tau \bar{d}(E)}
\end{equation}
where the density of states $d(E) = \sum_n \delta(E - E_n)$
is divided into a mean part $\bar{d}(E)$ and an oscillatory
part $d_{\text{o}}(E)$. For systems with time-reversal
symmetry, or more generally an anti-unitary symmetry,
it is expected that the form factor agrees in the semiclassical
limit ($\hbar \rightarrow 0$) with that of the
Gaussian Orthogonal Ensemble (GOE)
of random matrix theory which has the expansion
\begin{equation} \label{kgoe}
K^{\text{GOE}}(\tau) = 2 \tau - 2 \tau^2 + {\cal O}(\tau^3)
\quad \text{as} \quad \tau \rightarrow 0 \; .
\end{equation}

The semiclassical approximation for the form factor is 
obtained by inserting Gutzwiller's trace formula for the
density of states into (\ref{kdef}) and evaluating the
integral in leading order of $\hbar$. The result is an
approximation in terms of a double sum over all periodic
orbits of the classical system
\begin{equation} \label{ksemi}
K(\tau) \approx \frac{1}{h \bar{d}(E)} \left\langle
\sum_{\gamma, \gamma'} A_\gamma \, A^*_{\gamma'} 
\text{e}^{\frac{i}{\hbar} (S_\gamma - S_\gamma')} \, 
\delta\!\left(T\!-\!\frac{T_\gamma + T_{\gamma'}}{2}
\right)\!\right\rangle_E
\end{equation}
where $\tau = T/(h \bar{d}(E))$ and $h=2 \pi \hbar$. Furthermore,
$A_\gamma$ is an amplitude, generally complex-valued,
which depends on the stability and the Maslov index of the
periodic orbit $\gamma$, and $S_\gamma$ and $T_\gamma$ are
its action and period.

The double sum in (\ref{ksemi}) runs over all possible
pairings of periodic orbits. However, most of these pairs
do not contribute in the semiclassical limit.
Periodic orbits which are located in different regions
in phase space are uncorrelated, and when summed over,
the contributions from different pairs cancel each other.
It is expected that the relevant semiclassical contributions
come from a relatively small number of pairs of orbits
which are correlated. The key problem is then to identify
the mechanism which is behind these correlations.

The basic assumption we make is that
only those periodic orbits which are almost
everywhere close to one another, or to the time-reverse
of the other orbit, are correlated \cite{SR01}.
In order for two orbits to be different
but nevertheless close they must have special forms
which can be constructed in the following way.
The orbits are composed of different segments during
which one orbit follows very closely the other orbit
(or its time-reverse). However, the orbits can differ
in the way in which the segments are connected.
\begin{figure}
\begin{center}
\mbox{\epsfxsize12cm\epsfbox{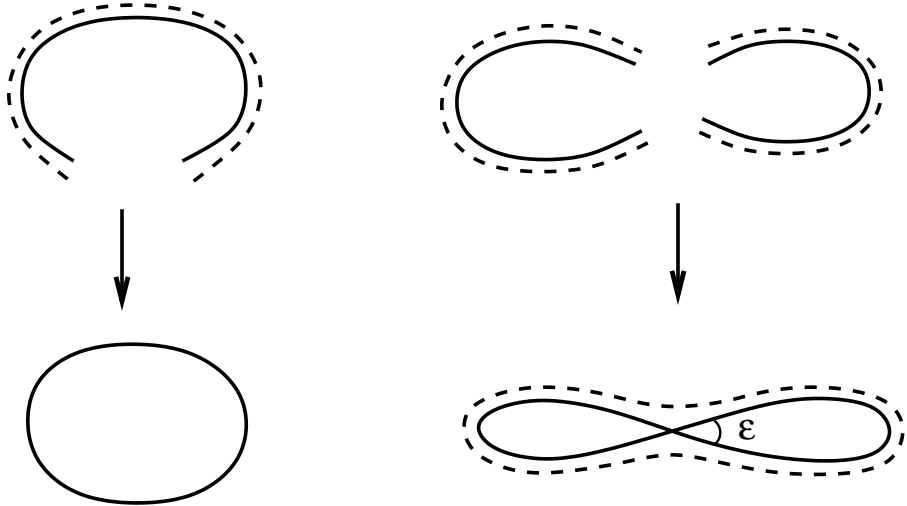}}
\end{center}
\caption{\label{figone}
The pairs of orbits considered here consist
of different segments. In each segment one orbit is very close
to the other (or its time reverse), but they can differ in
the way the segments are connected.}
\end{figure}

The two simplest possibilities are shown in Fig.~\ref{figone}.
If orbits are composed of only one segment then the two ends
can be connected in only one way. It then follows that
the two neighbouring orbits are either identical or one is
the time-reverse of the other. Including only these pairs
in the double sum corresponds to the diagonal approximation,
which yields the correct leading order behaviour $K(\tau)
\sim 2 \tau$ as $\tau \rightarrow 0$ \cite{Ber85}. 

Two segments, on the other hand, can be connected in two
ways, leading to orbits with or without self-intersection at the
connection point, as shown in Fig.~\ref{figone}.
In order for these pairs to exist and to be close, the crossing
angle $\ve$ has to be small. Then it can be shown in a linearized
approximation, that one orbit is indeed in the neighbourhood of the other.

In the following we evaluate the contributions of such
pairs of orbits to the spectral form factor. In order to
avoid further assumptions and to keep the calculations
simple we restrict attention now to systems with uniformly
hyperbolic dynamics, specifically we consider the representative
example of the geodesic motion on Riemann
surfaces with constant negative curvature \cite{BV86}.
Then the quantities $A_\gamma$, $S_\gamma$
and $T_\gamma$ in (\ref{ksemi}) depend only on the length
of an orbit, and $A_\gamma$ is positive.
We assume that the systems have no
further symmetries and are non-arithmetic so that the typical
degeneracy of a length of a periodic orbit is two.
For these systems the action difference for the
pairs of orbits being considered here
has been calculated in the linearized 
approximation for small crossing angle $\ve$ in \cite{SR01}
and is given by
\begin{equation} \label{dels}
\Delta S(\ve) \approx \frac{p^2 \ve^2}{2 m \lambda}
\end{equation}
where $\lambda$ is the Lyapunov exponent of the system,
and $p$ and $m$ are momentum and mass of the particle.

The sum over these pairs of orbits can be evaluated by
summing over all self-intersections of periodic orbits
with small crossing angle $\ve$, because for every such
self-intersection there exists a neighbouring periodic
orbit with action difference $\Delta S(\ve)$.
The self-intersections are determined by introducing a
function which selects them. This is done in the following
way. A self-intersection of a periodic orbit with
period $T$ divides the orbit into two loops. It can be
characterized by the crossing angle $\ve$ and the total
time $t$ along the shorter of the two loops, $t \leq T/2$.
Furthermore, we introduce an angle variable $\phi$ that specifies
the direction of the velocity, and a variable $t'$ that
measures the time along a periodic orbit.
If at any time $t'$ along a periodic orbit $\q(t'+t)=\q(t')$
and $\phi(t'+t) = \phi(t') - \pi + \ve$ then this periodic
orbit has a self-intersection with opening angle $\ve$, and
traversing the corresponding shorter loop takes time $t$.

Correspondingly, we can express the contribution from pairs
of the two-loop orbits to the form factor as
\begin{eqnarray} \label{k2a}
K^{(2)}(\tau) = & \frac{4}{h \bar{d}(E)} \Re
\int_{-\pi}^\pi \! d \ve \; \int_0^{T/2} \! d t \;
\sum_{\gamma} A_\gamma^2 \,
\text{e}^{\frac{i}{\hbar} \Delta S(\ve)}
\nonumber \\ & \times
\delta(T - T_\gamma) \, \int_0^T \! d t' \; f_{\ve,t}(\q(t'),\p(t'))
\end{eqnarray}
where the function $f_{\ve,t}$ is given by
\begin{eqnarray} \label{fun}
f_{\ve,t}(\q(t'),\p(t')) = & |J| \, \delta(\phi(t'+t) - \phi(t') +\pi - \ve)
\nonumber \\ & \times
\delta(\q(t'+t)-\q(t'))
\end{eqnarray}
Here $|J| = v^2 \, |\sin \ve| / \sqrt{g}$ is the Jacobian for
the transformation from the arguments of the three delta
functions to the three integration variables, where $v$ is
the speed of the particle and $g$ is the determinant of the
metric tensor. The three integrals give a contribution each
time that $t'$ is at the beginning of a loop with time $t$
and opening angle $\ve$. The choice of the limits of the
integral over $\ve$ is not important since the main contribution
in the semiclassical limit $\hbar \rightarrow 0$ comes from the
asymptotic behaviour of the integrand at $\ve = 0$.
In (\ref{k2a}) the amplitudes and the periods of the neighbouring
orbits were set to be equal since the difference does not
contribute to the leading semiclassical order.

One of the important properties of long periodic orbits
is their uniform distribution on the energy surface in phase
space. It implies that the average of a given phase space
function $f(\q,\p)$ along all periodic orbits of a certain
period $T$ can be replaced, in the limit $T \rightarrow \infty$,
by an average of this function over the energy surface in phase
space \cite{PP90}. More accurately, the following asymptotic
relation holds as $T \rightarrow \infty$
\begin{eqnarray} \label{uniform}
& \sum_{\gamma} |A_\gamma|^2
\delta(T - T_\gamma) \, \int_0^T \! d t' \; f(\q(t'),\p(t'))
\nonumber \\
\sim & \frac{T^2}{\Sigma(E)} \int \! d^2 q \, d^2 p \;
\delta\left(E - \frac{p^2}{2 m} \right) \; f(\q,\p)
\end{eqnarray}
where $\Sigma(E)$ is the volume of the energy surface in
phase space.

The relation (\ref{k2a}) is in the form in which this property of
the periodic orbits can be applied. The semiclassical limit
$\hbar \rightarrow 0$ is performed with the condition 
that $\tau / \hbar \rightarrow \infty$. The mean density
of states being $\bar{d}(E) \sim \Sigma(E) / (2 \pi \hbar)^2$,
this implies that $T \rightarrow \infty$
and thus the leading order semiclassical behaviour
arises from the large $T$ behaviour. Applying the uniformity
of the periodic orbit distribution and performing the integral
over the energy delta-function one obtains
\begin{equation} \label{k2b}
K^{(2)}(\tau) \sim \frac{4 p^2 T^2 }{m h \bar{d}(E)} \Re
\int_{-\pi}^\pi \! \! d \ve \; 
\text{e}^{\frac{i p^2 \ve^2}{2 m \hbar \lambda}} \sin|\ve| 
\int_0^{T/2} \! \! \! d t \,
p_E(\ve,t)
\end{equation}
where
\begin{equation} \label{probdens}
p_E(\ve,t) = \int \! \frac{d^2 q_0 \, d \phi_0}{\Sigma(E)}
\, \delta(\q(t) - \q_0) \, \delta(\phi(t) - \phi_0 + \pi - \ve)
\end{equation}
and $\q(t)$ and $\phi(t)$ are the coordinates of a particle
at time $t$, whose initial conditions at $t=0$ are specified
by $\q_0$, $\phi_0$ and energy $E$.

The quantity $p_E(\ve,t)$ has a direct classical interpretation.
It is the probability density for a particle with energy $E$ to
return after time $t$ to its starting point with a velocity
that deviates from the initial velocity by an angle $\ve - \pi$.
In the same way as for the diagonal approximation, one thus finds
that the periodic orbit sum is related to a transition
probability density in phase space \cite{AIS93}.

Our aim is to determine the leading order behaviour of (\ref{k2b})
as $\hbar \rightarrow 0$ which, as remarked above, depends on
the long-time behaviour of $p_E(\ve,t)$. For long times $p_E(\ve,t)$
approaches one over the volume of the energy shell in 
phase space, because the particle is equally likely to be found 
anywhere on the energy shell, i.e.\ 
$p_E(\ve,t) \sim 1/\Sigma(E)$ as $t \rightarrow \infty$.
Inserting this into (\ref{k2b}) and applying the method of
stationary phase yields
\begin{equation} \label{leading}
\frac{p^2 \tau^3 \Sigma(E)}{m \pi^2 \hbar^2} \Re
\int_0^\infty \! \! d \ve \; 
\exp \left( \frac{i p^2 \ve^2}{2 m \hbar \lambda} \right)
\; \ve = 0
\end{equation}
and so the leading order term as $\hbar \rightarrow 0$ vanishes.
This implies that one has to take into account the next order terms.
A closer analysis of (\ref{k2a}) shows that the important term
to consider is the next to leading order behaviour of $p_E(\ve,t)$
as $t \rightarrow \infty$. Quite surprisingly, the
two-loop contribution does not originate from the
ergodic limit of the probability density $p_E(\ve,t)$
but from the approach to this limit.
\begin{figure} 
\begin{center}
\mbox{\epsfxsize12cm\epsfbox{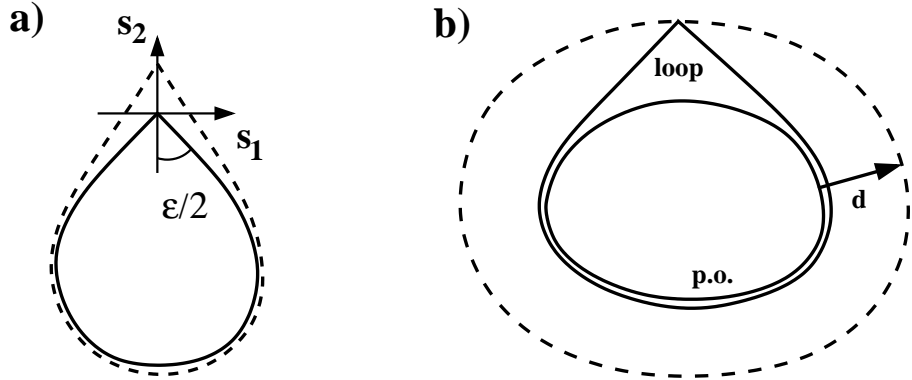}}
\end{center}
\caption{\label{figloop}
a) A loop with opening angle $\varepsilon$ (full line),
and the local coordinate system at its starting point.
b) Loops with the same opening
angle $\varepsilon$ whose traversal takes time $t$ form
a continuous family which have their starting points
on a curve of constant distance $d$ from a periodic orbit.}
\end{figure}

We have to consider $p_E(\ve,t)$ in
more detail. It is a classical transition probability
density and can be expressed in terms of classical
trajectories. These trajectories are all time $t$ loops
with opening angle $\ve$.
Consider one such loop as shown in Fig.~\ref{figloop}a.
Every point in the vicinity of its starting point is the starting
point of another loop, one example being shown by the dashed line.
To determine how angle $\ve$ and time $t$ change with 
the initial point we introduce a local coordinate system
(see Fig.~\ref{figloop}a) and linearize the motion
in the vicinity of the loop. The result is
\begin{equation} \label{linea}
v \, d t = 2 \cos \frac{\ve}{2}
\, d s_2 \, , \quad v \, d \ve = - 2 \lambda \,
\sin \frac{\ve}{2} \,
\tanh \frac{\lambda t}{2} \, d s_2
\end{equation}
One finds that angle and time change only in the $s_2$ direction,
but not in the $s_1$ direction. This is a particular property
of the uniformly hyperbolic dynamics. After integrating
the equations (\ref{linea}) one arrives at the following
conclusion. The loops with fixed $\ve$ and $t$ form continuous
one-parameter families. All the initial points of the loops
within a family lie on a curve which has a constant distance
(denoted by $d$) from a periodic orbit as shown schematically
in Fig.~\ref{figloop}b. The relation between the loops and
the periodic orbit is given by 
\begin{equation}
\cosh \frac{\lambda t}{2} \; \sin \frac{|\ve|}{2} =
\cosh \frac{\lambda t_0}{2}
\end{equation}
where $t_0$ is the period of the periodic orbit. It is a remarkable
property that any loop is uniquely related to a periodic orbit
into which it can be continuously deformed through a series of
other loops. Put another way, this
implies that any self-intersection of any arbitrary classical
trajectory is uniquely related to a periodic orbit, because a
self-intersection is the initial point of a loop.

We examined this property numerically. We chose a large
number of long random trajectories 
on a Riemann surface with constant negative curvature \cite{AS88}
and recorded all their
self-intersections. For every self-intersection a point is
plotted in the $(\ve,t)$-plane, where $\ve$ and $t$
are the opening angle and traversal time of the corresponding loop.
The result is shown in Fig.~3. As expected, the points form
continuous lines that start at the periods of the periodic
orbits (the $t$-values at $\ve=\pi$). One can observe a
logarithmic divergence of the curves at $\ve=0$ which is
implied by Eq.~(12). The full line in Fig.~3 is an evaluation
of Eq.~(12) for the second family of loops, and it is found to
be in perfect agreement with the numerical result.
\begin{figure} 
\begin{center}
\mbox{\epsfxsize12cm\epsfbox{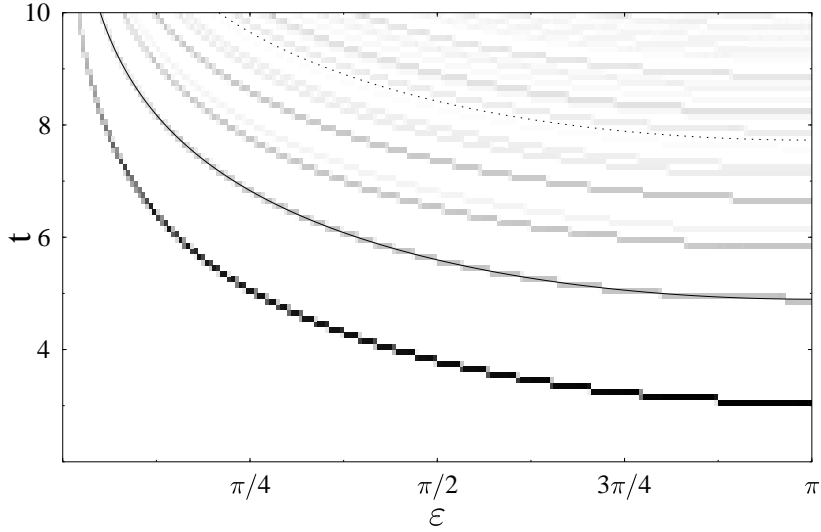}}
\end{center}
\caption{
Numerical result of the search for loops with
opening angle $\varepsilon$ and time $t$. Grey scales
are proportional to the number of loops found in bins
in the $(\ve,t)$-plane. 
}
\label{examp}
\end{figure}

We continue by expressing $p_E(\ve,t)$ in terms of the classical
trajectories. By evaluating the integrals over the
delta-functions in (\ref{probdens}), $p_E(\ve,t)$ can
be written as a sum over all families of loops with
opening angle $\ve$, which are labelled by $\xi$ in the
following. Alternatively, by using the relation (12),
$p_E(\ve,t)$ can also be expressed in terms of the
periodic orbits labelled by $\xi_0$.
\begin{eqnarray}
p_E(\ve,t) = & \frac{1}{\Sigma(E)}
\sum_{\xi} \frac{T_{\xi_0} \cosh(\lambda d/v) \,
\delta(t - T_\xi)}{\sin|\ve/2| \, (\Tr M_\xi - 2)} 
\nonumber \\
= & \frac{1}{\Sigma(E)}
\sum_{\xi_0} \frac{T_{\xi_0}  \, \delta(t - T_\xi)}{
\sqrt{(\Tr M_{\xi_0} - 2) (\Tr M_{\xi_0} - 2 + 4
\cos^2 \frac{\ve}{2})}} 
\end{eqnarray}
where $M_{\xi}$ and $M_{\xi_0}$ denote the stability matrices. 
We remark that a further use of Eq.~(12) yields
\begin{equation}
p_E(\ve,t) = p_E(\pi,t_0) \; .
\end{equation}
This means that the distribution $p(\ve,t)$ is identical to
the return probability density $p(\pi,t_0)$ at a shifted time
$t_0$, the relation between $t$ and $t_0$ being given by Eq.~(12).

Eq.~(13) is now applied to find the next
to leading order behaviour of the time integral over
$p_E(\ve,t)$ as $t \rightarrow \infty$. We assume that from
a certain time $T_0(\ve)$ on we can replace $p_E(\ve,t)$ by
its ergodic limit $(2 \pi m A)^{-1}$. This time $T_0(\ve)$ is
chosen to have the same $\ve$-dependence as the time of
the families of loops (like, for example, the dashed line in Fig.~3).
Thus $T_0(\ve)$ is related to $T_0(\pi)$
by an equation identical to that between $t$ and $t_0$ (Eq.~(12)).
For $t < T_0(\ve)$ we
replace $p_E(\ve,t)$ by its exact form, Eq.~(13). 
The approximation can be made asymptotically exact by letting
$T_0(\pi) \rightarrow \infty$ as $T \rightarrow \infty$. We find
\begin{eqnarray}
\int_0^{T/2} \! \! \! d t \; p_E(\ve,t) \sim &
\int_{T_0(\ve)}^{T/2} \! d t \; 
\frac{1}{\Sigma(E)} \; + \sum_{T_{\xi_0} < T_0(\pi)}
B_{\xi_0}(\ve)
\nonumber \\ = &
\frac{T/2 - T_0(\ve)}{\Sigma(E)} + \text{const} + {\cal O}(\ve^2)
\end{eqnarray}
where here and in the following constant denotes independence of $\ve$.
In the semiclassical limit only the asymptotic behaviour of Eq.~(15)
as $\ve \rightarrow 0$ is relevant and from the analog of Eq.~(12) we
find $ T_0(\ve) \sim - \frac{2}{\lambda} \log \ve + 
\text{const}$. This logarithmic divergence can be interpreted
as follows. For small
$\ve$ the two legs of a loop need a certain minimal time in
order to separate enough to enable the loop to close. This time
can be estimated by requiring that $\ve \exp(\lambda t/2)$
is of order one, yielding the logarithmic dependence above.
Substitution into (\ref{k2b}) results in
\begin{eqnarray}
K^{(2)}(\tau) \sim & \frac{8 p^2 T^2}{m h \bar{d}(E)} \Re
\int_0^\infty \! \! d \ve \;
\text{e}^{\frac{i p^2 \ve^2}{2 m \hbar \lambda}} \;
\frac{2 \ve (\log \ve + \text{const.})}{\lambda \Sigma(E)}
\nonumber \\
= & \frac{16 \tau^2}{\pi} \Re \int_0^\infty \! d \ve' \;
\text{e}^{i {\ve'}^2} \, \ve' \, \log(\ve') \, .
\end{eqnarray}
Evaluating the real part of the last integral finally
yields $K^{(2)}(\tau) \sim -2 \tau^2$ in agreement
with the $\tau^2$-term of the GOE form factor in (\ref{kgoe}).

In conclusion, we have shown that the off-diagonal contributions
to the spectral form factor from two-loop orbits yield a
$\tau^2$-term in agreement with random matrix theory. Its origin
can be traced to properties of loops with small opening angle
$\ve$. It is expected that higher-order terms in the expansion of $K(\tau)$
are related to multi-loop orbits, a point which is under investigation.

\end{document}